\documentclass[a4paper,11pt]{article}
\usepackage[parfill]{parskip}
\usepackage{physics, tensor, float, subcaption}
\usepackage{lineno}
\usepackage{jheppub}
\usepackage{amssymb,amsmath,amsthm}
\usepackage{mathrsfs}
\usepackage[utf8]{inputenc}
\usepackage{enumerate}
\usepackage{bigints}
\usepackage{float}
\usepackage{tikz}
\usepackage{setspace}
\usepackage{cancel}
\usepackage{array}
\usepackage{tabulary}
\usepackage{doi}
\usepackage{comment}

\definecolor{lime}{HTML}{A6CE39}
\newcommand{\orcidicon}{%
	\begin{tikzpicture}
	\draw[lime, fill=lime] (0,0) 
		circle [radius=0.16] 
		node[white] {{\fontfamily{qag}\selectfont \tiny ID}};
	\draw[white, fill=white] (-0.0625,0.095) 
		circle [radius=0.007];
	\end{tikzpicture}
	\hspace{-5mm}
}
\newcommand\orcidRudeep{{\href{https://orcid.org/0009-0002-0162-562X}{\orcidicon}}}

\protected\def\verythinspace{%
  \ifmmode
    \mskip0.5\thinmuskip
  \else
    \ifhmode
      \kern0.08334em
    \fi
  \fi
}

\newcommand{\thin}{\verythinspace}

\newcommand{\D}{\mathrm{d}}

\newcommand{\laund}[1]{\mathcal{O}{\left(#1\right)}}

\newcommand{\lie}[2]{\mathcal{L}_{\mathbf{\xi}}g_{#1\thin#2} = \xi^{\alpha}\partial_{\alpha}\thin g_{#1\thin #2} + g_{\alpha\thin #2}\thin\partial_{#1}\thin\xi^{\alpha}+g_{#1\thin \alpha}\thin\partial_{#2}\thin\xi^{\alpha}}

\begin{document}

\title{\vspace{-25pt}\huge{The Kerr Memory Effect at Null Infinity}}

\author[]{Rudeep Gaur\!\orcidRudeep}

\affiliation[]{School of Mathematics and Statistics, Victoria University of Wellington, 
\null PO Box 600, Wellington 6140, New Zealand.}

\emailAdd{rudeep.gaur@sms.vuw.ac.nz}
\abstract{We compute the memory effect due to a gravitational wave striking a Kerr black hole as seen by an observer at null infinity. This is done by working in Bondi--Sachs coordinates. It was shown by Hawking, Perry, and Strominger (HPS) that the memory effect due to a gravitational shockwave is seen as a pure BMS supertranslation from null infinity. Hence, it is of interest to compute the supertranslated Kerr solution in Bondi--Sachs coordinates. Finally, the gravitational wave is said to implant soft supertranslation hair on the event horizon of the black hole which carries superrotation charge. We will explicitly calculate the change in superrotation charge on the event horizon due to the supertranslation hair.
\vspace{1em}

\bigskip
\noindent
{\sc Date:} 11 March 2024; \LaTeX-ed \today

Memory effect;  Supertranslations.

\bigskip
\noindent{\sc} 

}

\maketitle

\clearpage
\section{Introduction} 
Since the observational discovery of gravitational waves nearly a decade ago by the Laser Interferometer Gravitational-Wave Observatory (LIGO) (and other detectors such as Virgo and KAGRA), many have wondered about detection of the gravitational memory effect --- the permanent alteration of a system due to a transient gravitational wave. The memory effect has been discussed in the literature since 1972, first introduced by Zel'dovich and  Polnarev \cite{gravmem1st}, then greatly expanded upon in the last few decades by Christodoulou and others \cite{chriswork1,Thorne:1992sdb,metricofsimpleexample,nonlineargraviationalwave,Tolish,tolishthesis,Garfinkle:2022dnm}. In the past few years there has been a deep mathematical connection made between the gravitational memory effect and a set of infinite symmetries at null infinity \cite{strominger2016gravitational,compere_lectures,stromingernotes,HPS,HPSrot}. These symmetries are associated to a set of transformations known as \textit{supertranslations} and \textit{superrotations}, collectively known as \textit{supertransformations}. In fact, when two particles are left permanently displaced by a gravitational wave, the initial and final states are related by a supertranslation.\\

This group of infinite symmetries has been known of for nearly 60 years, first introduced by Bondi, van der Burg, Metzner and Sachs \cite{bondi1960gravitational,Bondi:1962px,Sachs:1962wk,cit:sachs1} - known as the BMS group\footnote{Unfortunately, van der Burg's name is often forgotten.}. In recent years, there has been further research and development of the BMS algebra \cite{Barnich:2009se,Barnich_troes,aspectsofbms/cft} and the charges associated with supertranslations and superrotations. These charges have led to a hope of better understanding the `scattering problem' in general relativity \cite{strominger2016gravitational,stromingernotes}. Furthermore, it seems that charges associated with supertransformations may play an important role in resolving --- part of --- the information loss problem \cite{Strominger:2017aeh,HPSrot,HPS}. This is addressed by asserting that gravitational waves implant soft supertranslation hair on the event horizons of black holes. This hair is then evaporated off to null infinity, thereby preserving information from past null infinity through to future null infinity. Therefore, detection of the memory effect may offer a better understanding of abstract mathematical ideas and how they may be physically realised. Unfortunately, the memory effect will likely not be observationally detected until after the Laser Interferometer Space Antenna (LISA) \cite{Lisa,LISA:2022kgy} is launched.\\

Here we will discuss the effects of a transient gravitational shockwave striking the Kerr black hole as seen from an observer at futre null infinity\footnote{It is worth mentioning that we do not have an effective operational answer for ``where is null infinity'', nor do we claim to. It is an interesting question to ask, however. How does one define an operational notion of null infinity and how ``far one has to be from the source'' to observe something that is being radiated off to null infinity.}. We follow the calculations of HPS and others \cite{HPSrot,stromingernotes,cit:bhm} who have discussed the effects of a transient shock wave striking Schwarzschild and Reissner Nordstr\"om black holes. These authors show that the deformation of a black hole due to a gravitational shockwave (the memory effect) is seen as a pure BMS supertranslation \textit{from future null infinity}\footnote{There are additional effects that are not seen from null infinity, and this will be discussed in \autoref{memory_effect_and_ST}.}. Hence, we will compute the explicit supertranslated Kerr solution in Bondi--Sachs coordinates.\\

We will begin by summarising the expanded Bondi--Sachs metric, asymptotic Killing vectors which generate symmetries that have associated charges, and the relationship between the memory effect and supertranslations. In \autoref{Kerr_sec} we will discuss the Kerr solution in general Bondi--Sachs coordinates introduced by Fletcher and Lun \cite{fletcher2003kerr} and put it in the Bondi--Sachs gauge. In \autoref{sec_supertranslation} with the Kerr solution in the Bondi--Sachs gauge we will find the supertranslated metric functions. Finally, in \autoref{charges} we will calculate the supertransformation charges associated to the supertranslation hair that is implanted on the Kerr black hole due to the gravitational shockwave.

\subsection{The Bondi--Sachs Metric}

Bondi, van der Burg, Metzner, and Sachs \cite{bondi1960gravitational,Bondi:1962px,cit:sachs1} wanted to define a concept of asymptotic flatness at null infinity. The falloffs needed to be restrictive enough that unphyiscal spacetimes --- such as those with infinite energy --- would be ruled out, yet not so restrictive such that physical spacetimes and gravitational waves would be ruled out. While the falloffs may differ in the literature, we will use the choice made by BMS \cite{bondi1960gravitational,Bondi:1962px,cit:sachs1,stromingernotes}:
\begin{equation}\label{falloffs}
    \begin{aligned}
    &g_{uu}=-1 +\mathcal{O}(r^{-1}),\quad g_{ur} = -1 + \mathcal{O}(r^{-2}),\quad g_{uA}= \mathcal{O}(r^{0}),\\
    &g_{AB}= r^2\gamma_{AB} + \mathcal{O}(r),\quad g_{rr}=g_{rA}=0\,.
\end{aligned}
\end{equation}\\

The class of allowed asymptotic line elements for these falloffs is given by
\begin{equation}\label{generalBMsexpansion}
  \begin{aligned}
  ds^2 = &-\D u^2 -2\thin \D u\thin\D r + r^2\gamma_{AB}\thin \D \Theta^A \D \Theta^B\\ 
  &+\frac{2\thin m_{\,\text{bondi}}}{r}\D u^2 +r\thin C_{AB}\thin \D \Theta^A\D\Theta^B + D^BC_{AB}\thin\D u\thin \D\Theta^A\\
  &+\frac{1}{16\thin r^2}\thin \Big\{C^{FD}C_{FD}\Big\}\,\D u\thin \D r\\
  &+\frac{1}{r}\Big(\,\frac{4}{3}N_A +\frac{4u}{3}\partial_Am_{\,\text{bondi}}-\frac{1}{8}\partial_A\Big\{C^{FD}C_{FD}\Big\}\Big)\D u\thin \D\Theta^A\\
  &+\frac{1}{4}\gamma_{AB}\thin \Big\{C^{FD}C_{FD}\Big\}\,\D \Theta^A \D \Theta^B + \dots
\end{aligned}
\end{equation}
where $\Theta^A \in \{\theta,\phi\}$ and the uppercase Latin indices run over $\theta,\phi$. $D_A$ is the covariant derivative on the 2--sphere with respect to the 2--sphere metric, $\gamma_{AB}$. The function $m_{\text{bondi}}$ is the Bondi mass aspect, which is in general a function of $u$ and the angles, $\theta\,, \phi$. This can be used to obtain the entire Bondi mass after integrating $m_{\text{bondi}}$ over the entire 2--sphere at null infinity. In the case of the Kerr spacetime, the Bondi mass is simply, $M$, the mass of the black hole. $N_A$ is the angular moment aspect. Contracting $N_A$ with the generator of rotations and integrating over the entire sphere is related to the total angular momentum of the spacetime. $C_{AB}$ is another field which is symmetric and traceless ($\gamma^{AB}C_{AB} = 0$). The retarded time derivative of $C_{AB}$ is in fact the Bondi news tensor, 
\begin{equation}
    N_{AB} := \partial_{u}\thin C_{AB}\,.
\end{equation}
The news tensor is the gravitational analogue of the Maxwell field strength and its square is proportional to the energy flux across $\mathcal{I}^+$ \cite{stromingernotes,compere_lectures}.
\\

It is important to note that $N_{A}$ is defined slightly differently in various parts of the literature --- usually depending on the asymptotic expansion \eqref{generalBMsexpansion}. For instance, Comper\'e in refs \cite{Compere_final_state,45offinalstate} has a decomposition that leads to $N_A$ being defined as\footnote{This corresponds to pure vacuum and the mass contribution is omitted.}
\begin{equation}\label{barnichNAdef}
    N_A := -\frac{3}{32}\partial_{A}\Big(C_{BC}C^{BC}\Big) - \frac{1}{4}C_{AB}D_{C}C^{AC}.
\end{equation}
However, Strominger in ref \cite{stromingernotes} and Comper\'e in ref \cite{compere_lectures} uses the decomposition \eqref{generalBMsexpansion}, which leads to $N_A$ being defined as
\begin{equation}\label{stromingerNAdef}
    \frac{2}{3}N_A -\frac{1}{16}\partial_{A}\Big(C_{BC}C^{BC}\Big):= g^{(1)}{_{uA}}\,. 
\end{equation}
Here, $g^{(1)}_{uA}$ corresponds to the $r^{-1}$ expansion in $g_{uA}$. We will opt to use the second definition as the superrotation charge will not be changed. This can be seen explicitly in the case of the Kerr solution.\\

\subsection{Killing Vector}
Symmetries of a spacetime are associated to Killing vectors of that spacetime. Hence, before discussing the charges associated with symmetries in our spacetime, we must first briefly discuss the Killing vectors of our spacetime - or rather, the \textit{asymptotic Killing vectors}. The most general Killing vector, $\boldsymbol{\xi}$, that preserves the metric \eqref{generalBMsexpansion} to leading order is \cite{stromingernotes}
\begin{equation}\label{Full_Asymptotic_Killing_Vec}
    \begin{aligned}
        \xi^{\alpha}\partial_{\alpha} := f\partial_u &+ \Bigg[-\frac{1}{r}D^Af + \frac{1}{2\thin r^2}C^{AB}D_{B}f + \laund{\frac{1}{r^3}}\Bigg]\,\partial_A\\
        &+\Bigg[\frac{1}{2}D^2f -\frac{1}{r}\Bigg\{\frac{1}{2}D_Af D_B\thin C^{AB} + \frac{1}{4}C^{AB}\thin D_{A}D_B \thin f\Bigg\}+ \laund{\frac{1}{r^2}} \Bigg]\,\partial_r\,.
    \end{aligned}
\end{equation}
Here, $f$ is a function of that angular coordinates $(\theta,\phi)$ only and $D^2$ is the standard Laplacian on the 2--sphere. However, since the analysis conducted here is only a calculation to linear order --- as done by HPS \cite{HPSrot}, the Killing vector is truncated:
\begin{equation}\label{Asymptotic_Killing_vec_truncated}
    \xi^{\alpha}\partial_{\alpha} = f\thin \partial_u + \frac{1}{2}D^2 f \partial_r - \frac{1}{r}D^Af\partial_A\,.
\end{equation}\\
This does indeed beg the question of whether a second order analysis would still show that the memory effect --- as seen from null infinity --- is still a supertranslation with the Killing vector \eqref{Full_Asymptotic_Killing_Vec}.

\subsection{Associated Charges and Charge Conservation}
It is well known that the symmetries of spacetime are associated to conserved charges via Noether's theorem. Before the discovery of the BMS group, the largest symmetry group was the Poincar\'e group, which has associated conserved charges such as energy and momentum. The BMS group, which is an infinite dimensional group of symmetries at null infinity also has charges associated to supertranslations and superrotations.\\

Supertranslation charge and its conservation is given by\footnote{Note that integration is carried out over the 2--sphere at null infinity.} \cite{stromingernotes}
\begin{equation}\label{supertranslation_charge}
    Q_f^+ = \frac{1}{4\pi} \int_{\mathcal{I}^+} \D^2 \Theta \,\sqrt{\gamma}\,fm_{\,\text{bondi}} =  \frac{1}{4\pi} \int_{\mathcal{I}^-} \D^2 \Theta\, \sqrt{\gamma}\,fm_{\,\text{bondi}} = Q^-_f\,.
\end{equation}
Here $f$ is a function of angular coordinates\footnote{Note that while $f$ is often chosen to be a spherical harmonic function, it is not limited to be only a spherical harmonic.} and can be thought of as the generator of supertranslations. In general, the supertranslation charges will depends on advanced/retarded time. This is simply due to $m_{\text{bondi}}$ --- in general --- depending on advanced/retarded time. In the case where $f=1$ the conserved charge is energy while in the case where $f$ is a $l=1$ spherical harmonic function we have conservation of ADM momentum.\\

Superrotation charge and its conservation is given by\footnote{This definition of superrotation charge follows from using \eqref{stromingerNAdef} as our definition of $N_A$. In the case that one uses \eqref{barnichNAdef}, superrotation charge is expressed via \cite{Barnich_troes} 
\begin{equation*}
     Q_Y = \frac{1}{16\pi}\int \D ^2\Theta \,\sqrt{\gamma}\, Y^A\Big[ 2N_A +\frac{1}{16}\partial_{A}\Big(C_{BC}C^{BC}\Big)\Big]\,.
\end{equation*}}
\begin{equation}\label{superrotation_charge}
    Q^+_Y = \frac{1}{8\pi}\int_{\mathcal{I}^+} \D^2\Theta \,\sqrt{\gamma}\, Y^A N_A = \frac{1}{8\pi}\int_{\mathcal{I}^-} \D^2\Theta\, \sqrt{\gamma}\, Y^A N_A = Q^-_Y,
\end{equation}
where $Y^{A}$ is an arbitrary vector field on the 2--sphere \footnote{As is the case for $f$, the \textit{components} of $Y^A$ are often chosen to be spherical harmonic functions, however, the components of $Y^A$ are not limited to be spherical harmonics.}. In the case that $Y^A$ is one of the 6 global conformal Killing vectors on the 2--sphere \eqref{superrotation_charge} expresses conservation of ADM angular momentum and boost charges.\\

The notation used here indicates that \textit{supertransformation} charge at future null infinity, $\mathcal{I}^+$ should \textit{match} the supertransformation charge at past null infinity, $\mathcal{I}^-$. Furthermore, these \textit{matching conditions} are a statement about having a well-posed scattering problem in general relativity. The existence of these conserved charges is --- at least in principle --- verifiable with the gravitational memory effect, which may be detectable in the near future.

\clearpage

\subsection{The Memory Effect and Supertranslations}\label{memory_effect_and_ST}

During the last decade there has been a lot of discussion regarding the gravitational memory effect and the direct correspondence of this effect with the BMS group --- for instance, one may see refs \cite{stromingernotes,HPSrot,HPS,compere_lectures,Compere_final_state,donnaysuperathorizon,extendedsymm,Wald_bHM}. It was shown by HPS \cite{HPSrot} that a deformation due to a shockwave defined via an energy-momentum tensor is equivalent to a BMS supertranslation at null infinity (for a Schwarzschild black hole). Such a supertranslation is given by taking the Lie derivative of the spacetime metric --- in this case, the Schwarzschild metric --- along the asymptotic Killing vector\footnote{Note the sign change here due to changing from retarded time to advanced time.}, 
\begin{equation}\label{asymptoticKillingVector}
    \xi^{\mu} \partial_{\mu} = f\partial_{v} - \frac{1}{2}D^2 f\partial_r + \frac{1}{r}D^{A}f\partial_A\,.
\end{equation}
However, as HPS \cite{HPSrot} state, supertranslations only equate to part of the deformation a gravitational wave would produce when striking a black hole.\\

Intuitively, gravitational waves, which carry energy, should impart some of this energy to the black hole it strikes and it should alter the \textit{mass} and/or (angular) momentum. In fact, it was shown by HPS \cite{HPSrot} that there was a change in the mass of a Schwarzschild black hole due to the monopole (denoted by $\mu$ in  ref \cite{HPSrot}) contribution of the shockwave. Therefore, the full\footnote{The word \textit{full} is used here, however, it is worth noting that this is only an analysis at linear order.} ``memory effect" due to such a shockwave was written as\footnote{The Heaviside terms here have been neglected as we are assuming the shockwave has already struck the black hole.}
\begin{equation}\label{real_change}
    \delta g_{\mu\nu} = \mathcal{L}_{\xi} g_{\mu\nu} + \frac{2\mu}{r}\delta^{\thin v}{_{\mu}}\delta^{\thin v}{_{\nu}}\,.
\end{equation}
Here $\delta g_{\mu\nu}$ refers to the permanent change in the spacetime due to a gravitational wave. Therefore, the mass of the hairy Schwarzschild black hole would be $m = M + \mu$ after the gravitational shockwave strikes the black hole. However, while the mass of the black hole may change, the \textit{Bondi mass} of the black hole --- at least at linear order --- does not change. This further emphasises that the memory effect is not \textit{entirely captured by BMS supertransformations}. However, the memory effect as seen from null infinity \textit{is entirely captured by BMS supertranslations}.\\

This discussion is important when analysing the calculations in this paper. Applying the same gravitational shockwave used in \cite{HPSrot,stromingernotes,cit:bhm} to the Kerr solution is a non-trivial task and thus one does not know for certain what the other components to the memory effect will appear. A common suspicion within the community is that there may be another monopole contribution to the mass term that appears in the $g_{u\phi}$ component of the metric. This would imply a change in the angular momentum of the Kerr black hole due to a gravitational shockwave --- which will not be seen at null infinity by observation of the superrotation charge.\\

\clearpage

\section{The Kerr Metric}\label{Kerr_sec}
We will now discuss the Kerr metric in general Bondi--Sachs coordinates. The Kerr solution was first introduced in generalised Bondi--Sachs coordinates by Fletcher and Lun \cite{fletcher2003kerr} as they were interested in investigating gravitational radiation in the Kerr spacetime.  
As we are interested in the behaviour of this solution at large $r$, i.e., expanded in powers of $1/r$ we shall only present the asymptotic line element\footnote{We have chosen to not list the $g_{\theta\phi}$ term here since it is of order $\laund{1/r}$. Comparing this to \eqref{generalBMsexpansion} it is clear that this term will be of subleading order and not used in the analysis.} found in the appendix of ref \cite{fletcher2003kerr}:
\begin{equation}\label{kerrbondicoord}
  \begin{aligned}
    \D s^2 =& -\Bigg(1-\frac{2M}{\Tilde{r}
}\Bigg)\,\D u^2 -2\Bigg(1+\frac{a^2 \cos^2\theta-\frac{1}{2}\thin a^2}{\Tilde{r}
^2}\Bigg)\thin \D u\thin \D \Tilde{r}
 \\
    &-2\cos\theta \Bigg(a -\frac{2aM + 2a^2 \sin\theta}{\Tilde{r}
} \Bigg)\thin \D u \thin \D\theta -2\Bigg(\frac{2aM \sin^2\theta}{\Tilde{r}
}\Bigg)\thin \D u\thin \D\phi\\
    &+\Tilde{r}
^2 \Bigg(1 + \frac{2a\sin\theta}{\Tilde{r}
} + \frac{2a^2-3a^2\cos^2\theta}{\Tilde{r}
^2}\Bigg)\thin \D \theta^2\\
    &+\Tilde{r}
^2 \Bigg( \sin^2\theta - \frac{2a\sin\theta\cos^2\theta}{\Tilde{r}
} + \frac{a^2 - 3a^2\cos^2\theta + 3a^2\cos^4\theta}{\Tilde{r}
^2}\Bigg)\thin \D \phi^2 +\dots\\
  \end{aligned}
\end{equation}

The line element given in \eqref{kerrbondicoord} does not, however, satisfy the Bondi-Sachs gauge. This gauge is reached by requiring the coordinate, $r$, to be the ``luminosity distance'' \cite{bondi1960gravitational}. This is, however, misleading as in the term luminosity distance means something quite different in cosmology. In fact, $r$, is simply chosen to be an areal coordinate that varies along null rays that satisfies $\mathrm{det}(g_{AB}) = r^4 \sin^2\theta$. This is achieved by defining\footnote{In fact, this is not the only reason this transformation is chosen. This coordinate transformation also leads to the trace of $C_{AB}$ vanishing and to the metric taking the form \eqref{generalBMsexpansion}.},
\begin{equation}\label{redefinedr}
    \Tilde{r} := r + \frac{a}{2}\frac{\cos2 \theta}{\sin\theta} + \frac{a^2}{8}\,\Bigg(4\cos 2 \theta +\frac{1}{\sin^2\theta}\Bigg) \frac{1}{r}\,.
\end{equation}
With this radial coordinate the line element \eqref{kerrbondicoord} may be recast into the following form\footnote{The $g_{ur}$ component here differs from ref \cite{Barnich_troes}. This is in fact a small error in Appendix D of their paper and can be verified by computing $\D \Tilde{r}$. Furthermore, one may note that this component now correctly provides the scalar, $C^{AB}C_{AB}$.}:
\begin{equation}\label{Kerrbondigauge}
    \begin{aligned}
        \D s^2 =& -\Bigg(1-\frac{2M}{{r}}\Bigg)\,\D u^2 -2\Bigg(1-\frac{a^2}{16\thin\sin^2\theta}\frac{1}{r^2}\Bigg)\thin \D u\thin \D {r}\\
        &-2\Bigg(-\frac{a}{2}\frac{\cos\theta}{\sin^2\theta} -\frac{a\cos\theta}{4}\{8M + \frac{a}{\sin^3\theta}\}\frac{1}{r}\Bigg)\thin\D u\thin\D\theta -2\Bigg(\frac{2aM \sin^2\theta}{r} \Bigg)\thin \D u\thin \D\phi\\
        & + \Bigg(r^2 + \frac{a}{\sin\theta}\thin r + \frac{a^2}{2\thin\sin^2\theta}\Bigg)\thin \D \theta^2 +\Bigg( r^2\sin^2\theta -a\sin\theta\, r +\frac{a^2}{2}\Bigg)\thin \D \phi^2 +\dots
    \end{aligned}
\end{equation}
By comparing the expanded form of Kerr solution in the Bondi-Sachs gauge in \eqref{Kerrbondigauge}
with \eqref{generalBMsexpansion} 
we may read off $N_{A}$, $C_{AB}$, and $C_{AB}\thin C^{AB}$ for the Kerr solution.
\begin{equation}\label{various_components}
  \begin{aligned}
    &C_{AB}\thin C^{\thin AB} \equiv \frac{2a^2}{\sin^2\theta};\\
    &C_{AB}\thin \D x^{A}\thin \D x^{B} \equiv \frac{a}{\sin\theta}\,\D\theta^2 - a\sin\theta\, \D \phi^2; \\
    &N_{\theta} \equiv  3 a M\cos\theta \,;\\
    &N_{\phi} \equiv -3 a M \sin^2\theta;\\
    & m_{\,\text{bondi}} \equiv M\,.\\
  \end{aligned}
\end{equation}

One may note here that the coordinate transformation \eqref{redefinedr} is singular when $\sin\theta = 0$. However, this is in fact only a coordinate singularity as one may confirm by checking both the Ricci and Kretschmann scalars. Both of these scalar invariants remain finite when $\sin\theta = 0$. Intuitively, one \textit{may} see this as a result of using simple null geodesics --- those with zero angular momentum about the \textit{axis of symmetry} --- to arrive at this particular version of the Kerr solution. Simple null geodesics were used by Fletcher and Lun \cite{fletcher2003kerr} as the principal null directions of the Kerr solution do not form constant $u$ hypersurfaces. This feature has been discussed in the literature and does limit the applicability of this metric in numerical studies --- one may see refs \cite{Venter:2005cs,kdsinbondi} for further discussions.\\

Furthermore, note that we are using retarded time instead of advanced time. The original calculations by HPS \cite{HPSrot,stromingernotes} were done in advanced time. However, we are interested in the potential observation of the supertranslated Kerr black hole in the \textit{future} of the black hole being struck by a gravitational wave\footnote{By following the calculations of Fletcher and Lun \cite{fletcher2003kerr} changing from retarded time to advanced time will only require a sign change. Therefore, if one wishes to see charges \textit{implanted} on the horizon, one can easily do so by changing the a sign in the relevant components.}.\\

\section{Supertranslations of the Kerr Spacetime}\label{sec_supertranslation}
Unlike the cases considered in refs \cite{HPSrot,cit:bhm,KKSofthair} --- which are non-rotating spacetimes --- there are several other terms involved when calculating the diffeomorphisms of the Kerr solution in the form \eqref{kerrbondicoord}. All of these diffeomorphisms are explicitly given by
\begin{equation}
    \delta g_{\mu\nu} := \lie{\mu}{\nu}\,,
\end{equation}
where the $\boldsymbol{\xi}$ is once again the asymptotic Killing vector in \eqref{asymptoticKillingVector}.
The supertranslated metric functions are found to be:
\begin{equation}\label{supertranslated_kerr_functions}
    \begin{aligned}
        &\delta g_{uu} = -M\frac{D^2f}{r^2}\,; \\
        &\delta g_{ur} =\frac{1}{r^2}\frac{\thin a\cos\theta}{2\thin\sin^2\theta}\thin D^2f\,;\\
        &\delta g_{u\thin\theta} =-\Bigg(\partial_{\theta}f +\frac{1}{2}\partial_{\theta}D^2f\Bigg) +\frac{1}{r}\Bigg(2M\partial_{\theta}f - \partial_{\theta} \thin\Bigg\{\frac{a}{2}\frac{\cos{\theta}}{\sin^2\theta}\partial_{\theta}f\Bigg\}\Bigg)\,;\\
        &\delta g_{u\thin\phi} =  -\Bigg(\partial_{\phi}f + \frac{1}{2}\partial_{\phi}D^2f\Bigg)+ \frac{1}{r}\Bigg(2M\partial_{\phi}f - \frac{a}{2}\frac{\cos{\theta}}{\sin^2\theta}\partial_{\phi}\thin\partial_{\theta}f\Bigg)\,;\\
        &\delta g_{\theta\theta} =  \Bigg\{2\partial_{\theta}{^2}f - D^2f\Bigg\}\thin r - \frac{a}{\sin\theta}\Bigg\{+\frac{1}{2}D^2f + 2\thin\frac{\cos\theta}{\sin\theta}\partial_{\theta}f - 2\thin\partial_{\theta}{^2}f\Bigg\}\,;\\
        &\delta g_{\phi\phi}= \Bigg\{ 2\partial_{\phi}{^2}f +2\thin\sin\theta\cos\theta\,\partial_{\phi}f- \sin^2\theta D^2f\Bigg\}\thin r\\
        &\hspace{14.2em}-\frac{a}{\sin\theta}\Bigg\{\frac{1}{2}\sin^2\theta D^2f + \cos\theta\thin\partial_{\phi}f + 2\thin\partial_{\phi}{^2}f\Bigg\}\,.\\
    \end{aligned}
\end{equation}
Comparing the supertranslated metric functions with refs \cite{HPSrot,cit:bhm,KKSofthair} it is clear that upon setting $a=0$ we recover the supertranslated Schwarzschild black hole\footnote{The notation used in this paper is more explicit when comparing to the referenced papers. This leads to the $g_{\phi\phi}$ supertranslation looking slightly different. This is, however, due to the fact that the covariant derivatives have been calculated explicitly and not left in the form $D_A(D_Bf)$\,.}. One may note that the supertranslated $C_{AB}$ field is the same for the Kerr solution and Schwarzschild. This is to be expected as in refs \cite{stromingernotes,compere_lectures} is it shown that there should be no change unless $m_{\text{bondi}}$ is actually a function of retarded/advanced time.\\

The new metric functions are defined via
\begin{equation}
    \Bar{g}_{\mu\nu} := g_{\mu\nu} + \delta g_{\mu\nu}\,,
\end{equation}
which are:
\begin{equation}\label{supertranslated_kerr_metriccomps}
    \begin{aligned}
        &\Bar{g}_{uu} = -\Bigg(1-\frac{2M}{r}-M\frac{D^2f}{r^2}\Bigg)\\
        &\Bar{g}_{ur} = -\Bigg(1-\Bigg\{\frac{a^2}{16\thin\sin^2\theta} +\frac{8\thin a\cos\theta}{16\thin\sin^2\theta}D^2f\Bigg\}\frac{1}{r^2}\Bigg)\\
        &\Bar{g}_{u\theta} = -\Bigg\{-\frac{a}{2}\frac{\cos\theta}{\sin^2\theta} +\partial_{\theta}f + \frac{1}{2}\partial_{\theta}D^2f\Bigg\}\\
        &\hspace{2.2em}- \Bigg\{- \frac{a\cos\theta}{4}\{8M + \frac{a}{\sin^3\theta}\} -2M\partial_{\theta}f +\partial_{\theta} \Bigg[\frac{a}{2}\frac{\cos{\theta}}{\sin^2\theta}\partial_{\theta}f\Bigg]\Bigg\}\frac{1}{r}\\
        &\Bar{g}_{u\phi} = -\Bigg(\partial_{\phi}f + \frac{1}{2}\partial_{\phi}D^2f\Bigg) + \Bigg\{-2aM \sin^2\theta + 2M\partial_{\phi}f - \frac{a}{2}\frac{\cos{\theta}}{\sin^2\theta}\partial_{\phi}\thin\partial_{\theta}f\Bigg\}\frac{1}{r}\\
        &\Bar{g}_{\theta\theta} = r^2 + \Bigg\{\frac{a}{\sin\theta} + 2\partial_{\theta}{^2}f - D^2f\Bigg\}\thin r \\
        &\hspace{1.3em} - \frac{a}{\sin\theta}\Bigg\{-\frac{a}{2\sin\theta}+\frac{1}{2}D^2f + 2\thin\frac{\cos\theta}{\sin\theta}\partial_{\theta}f - 2\thin\partial_{\theta}{^2}f\Bigg\}\\
        &\Bar{g}_{\phi\phi} = r^2\sin^2\theta + \Bigg\{-a\sin\theta + 2\partial_{\phi}{^2}f +2\thin\cos\theta\sin\theta\,\partial_{\phi}f- \sin^2\theta D^2f\Bigg\}\thin r \\
        &\hspace{4em}- \frac{a}{\sin\theta}\Bigg\{-\frac{a\sin\theta}{2}+\frac{1}{2}\sin^2\theta D^2f + \cos\theta\thin\partial_{\phi}f + 2\thin\partial_{\phi}{^2}f\Bigg\}\,.
    \end{aligned}
\end{equation}
The supertranslated Kerr metric in \autoref{sec_supertranslation} is referred to as a ``hairy black hole''. The hair carried is soft supertranslation hair, which have corresponding charges. 
From the metric components, one can read off $C_{AB}C^{AB}$, $C_{AB}$, and $N_{A}$ after the Kerr spacetime has been supertranslated. 
\begin{equation}\label{supertranslated_various_comps}
    \begin{aligned}
       & C_{AB}\thin C^{AB} = \frac{2\thin a^2}{\sin^2\theta} + \frac{16\thin a\cos\theta}{\sin^2\theta}D^2f\\
       &C_{AB}\thin \D x^{A}\thin \D x^{B} = \Bigg(\frac{a}{\sin\theta} + 2\partial_{\theta}{^2}f - D^2f\Bigg)\,\D\theta^2 \\
       &\hspace{5.7em}-\Bigg( a\sin\theta - 2\partial_{\phi}{^2}f -2\thin\frac{\cos\theta}{\sin\theta}\partial_{\phi}f- \sin^2\theta D^2f\Bigg)\, \D \phi^2\\
       & N_{\theta} = 3M\{\thin a\cos\theta + \partial_{\theta}f\thin \} + \frac{3}{2}\thin a\thin\partial_{\theta}\Bigg\{\frac{\cos\theta}{\sin^2\theta}\Big[D^2f - \frac{1}{2}\partial_{\theta}f\Big] \Bigg\}\\
       &N_{\phi} = 3M\{\thin- a\sin^2\theta + \partial_{\phi}f\thin \} + \frac{3}{2}\thin a\thin\partial_{\phi}\Bigg\{\frac{\cos\theta}{\sin^2\theta}\Big[D^2f - \frac{1}{2}\partial_{\theta}f\Big] \Bigg\}\,,
    \end{aligned}
\end{equation}
and the supertranslated event horizon is now located at
\begin{equation}\label{shifted_horizon}
    (r_+)_f = r_+ +\frac{1}{2}D^2f\,.
\end{equation}
It is apparent that several terms in the supertranslated metric functions are once again singular when $\sin\theta=0$. If one is to use this solution for numerical studies then it is likely that $f$ must be restricted to very particular functions in order to eliminate this issue.

\section{Charges}\label{charges}
With the supertranslated spacetime --- the hairy Kerr black hole --- we may now discuss the superrotation charges associated to the supertranslation hair which are evaporated off the event horizon to future null infinity.\\

\subsection{Supertranslation Charge}

As discussed by HPS \cite{HPSrot,stromingernotes} supertranslation hair do not impart supertranslation charge. Recall that the supertranslation charge observed at future null infinity is defined as follows:
\begin{equation}
     Q_f = \frac{1}{4\pi} \int_{\mathcal{I}^+} \D^2 \Theta \,\sqrt{\gamma}\,fm_{\,\text{bondi}}\,.
\end{equation}
It is now clear that since the Bondi mass aspect, $m_{\text{bondi}}$, is not changed due to the \textit{supertranslation alone} we will not have any supertranslation charge turned on by supertranslation hair. 
However it is worth noting that in Donnay et al. \cite{cit:bhm} and ref \cite{KKSofthair} an analysis of the Schwarzschild and Kaluza--Klein spacetimes in the near horizon limit is conducted. This analysis shows that nontrivial supertranslation charge is turned on at the horizon due to the gravitational wave that is absent at null infinity. Furthermore, one may see 
ref \cite{nearhorizon_req} for an analysis relating near horizon displacement effects to supertransformation charges.\\ 

\subsection{Superrotation Charge}
Supertranslation hair does however, carry superrotation charge. The superrotation charge that is measured at future null infinity is given by
\begin{equation}
    Q_Y = \frac{1}{8\pi}\int_{\mathcal{I}^+} \D^2\Theta \,\sqrt{\gamma}\, Y^A N_A\,.
\end{equation}
Using \eqref{supertranslated_various_comps} we see that superrotation charge that is present at null infinity is
\begin{equation}\label{Superrotation_theta_charge}
    \begin{aligned}
    Q_{Y=Y^{\theta}} = \, &\frac{1}{8\pi}\int_{\mathcal{I}^+}  \sqrt{\gamma}\,\D^2\Theta\,Y^{\theta}\,3M a\cos\theta\,+\\
    & \frac{1}{8\pi}\int_{\mathcal{I}^+}\sqrt{\gamma}\,\D^2\Theta \,Y^{\theta}\Bigg[\partial_{\theta}f + \frac{3}{2}\thin a\thin\partial_{\theta}\Bigg\{\frac{\cos\theta}{\sin^2\theta}\Big[D^2f - \frac{1}{2}\partial_{\theta}f\Big] \Bigg\}\Bigg]\,,
    \end{aligned}
\end{equation}
and 
\begin{equation}\label{superrotation_phi_charge}
    \begin{aligned}
        Q_{Y=Y^{\phi}} =\,  &\frac{1}{8\pi}\int_{\mathcal{I}^+}-\sqrt{\gamma}\,\D^2\Theta\,Y^{\phi}\,3M a\sin^2\theta\, + \\
         &  \frac{1}{8\pi}\int_{\mathcal{I}^+}\sqrt{\gamma}\,\D^2\Theta \,Y^{\phi}\Bigg[\partial_{\phi}f + \frac{3}{2}\thin a\thin\partial_{\phi}\Bigg\{\frac{\cos\theta}{\sin^2\theta}\Big[D^2f - \frac{1}{2}\partial_{\phi}f\Big] \Bigg\}\Bigg]\,.
    \end{aligned}
\end{equation}
Here the first lines correspond to the bald Kerr black hole superrotation charges and can easily be recovered if the supertranslation function $f$ vanishes. Furthermore, as one would expect when $a=0$ we recover the superrotation charges of the hairy Schwarzschild black hole \cite{HPSrot,cit:bhm,KKSofthair}. As shown by Barnich and Troessaert in \cite{Barnich_troes}, when $Y^{\phi}$ is the Killing vector, $\partial_{\phi}$, \eqref{superrotation_phi_charge} corresponds to conservation of angular momentum (for the bald Kerr black hole). For the Hairy Kerr black hole, one may see that the zero-mode superrotation charge (when $f=0$ and $Y^{\phi} = 1$) given by \eqref{superrotation_phi_charge}, does not change and will still correspond to conservation of angular momentum.\\

Detection of these charges still remains an open question. One first requires an \textit{operational} notion of finite infinity. Secondly, we require a notion of what higher order charges would be observed as in our detectors. In theory, the supertranslation field should be detectable via classical tests of general relativity such as the bending of light \cite{Compere_final_state,Teo_lensing,stdetection_req}. However, as Comper\'e discusses in \cite{Compere_final_state}, one would need an array of detectors surrounding the central object in order to deduce the superrotation charges, thereby confirming the existence of the supertranslation field.\\

\subsection{Supertransformation charges and the Memory effect}

In \autoref{memory_effect_and_ST} we discussed that it has been shown the mass, $M$ --- in the case of the Schwarzschild spacetime --- is changed by a factor of $\mu$ which is the monopole contribution to the shockwave. We discussed that in the case of Kerr, this shockwave may also change the mass term present in the $g_{uA}$ terms. This would change the angular momentum of the Kerr black hole, as one may expect from the passing of a gravitational wave. 
Since the mass of the black hole is changed, one may ask why there is no supertranslation charge found at null infinity or why the zero-mode of superrotation charge (angular momentum) is not changed.\\

Recall, however, that supertranslation and superrotation charges are only defined for $\mathcal{I}^+$ and $\mathcal{I}^-$. This, therefore, becomes a statement of what an observer at null infinity measures as the memory effect rather than what the memory effect may be for \textit{all observers}. Indeed, it seems that the change in mass and, therefore, changes in momentum and/or angular momentum are not measurable at null infinity via the measurement of superrotation charges.

\clearpage

\section{Conclusion}\label{Conclusion}

We have studied the effects of a BMS supertranslation on the Kerr black hole in Bondi coordinates. This was first done by taking the asymptotic expansion of this solution which matches perfectly with the general BMS expansion of a asymptotically flat metric \eqref{generalBMsexpansion} after a coordinate change given by \eqref{redefinedr}. In \autoref{sec_supertranslation} we then found the supertranslated metric functions --- the hairy Kerr black hole --- which were used to to determine the supertranslation and superrotation charges that may be found at null infinity in \autoref{charges}.\\

We discussed the change in these supertransformation charges due to the supertranslation hair implanted on the Kerr black hole by the gravitational wave. It was shown that supertranslation charge was absent at null infinity. However, the supertranslation hair did in fact turn on superrotation charge that was detectable at null infinity, given by \eqref{Superrotation_theta_charge} and \eqref{superrotation_phi_charge}. We showed that the zero-mode of the superrotation charge remained unchanged at null infinity since any change in mass is not due to pure supertranslations. While detection of these charges still requires further technological developments, calculations of these charges does hope to provide a better understanding of the scattering problem when astrophysical, rotating black holes are involved.\\ 

The near horizon limit of the extremal Kerr solution was not discussed. It was shown in refs \cite{cit:bhm,KKSofthair} that there was a non-trivial supertranslation charge turned on at the horizon for the hairy Schwarzschild and Kaluza--Klien spacetimes. Hence, the near horizon Kerr limit would be interesting to explore. Furthermore, the charged Kerr black hole --- the Kerr--Newman solution would be interesting to explore as it has been shown that the presence of a vector potential will lead to \textit{soft electric hair} \cite{HPS}.

\acknowledgments
RG was supported by a Victoria University of Wellington PhD Doctoral Scholarship. RG would like to acknowledge Prof. Matt Visser for various comments on this paper along with Marco Galoppo, and Chris Harvey-Hawes for deep and interesting conversations regarding the infrared triangle. 

\clearpage


\clearpage

\bibliographystyle{JHEP}
\bibliography{Kerr_mem.bib}


\end{document}